\documentclass{aa}
\usepackage{graphicx}
\usepackage{txfonts}
\usepackage{natbib}

\begin{document}

\title{The ultra-cool white dwarf companion of PSR~J0751+1807}

\author{C. G. Bassa\inst{1}
   \and M. H. van Kerkwijk\inst{2}
   \and S. R. Kulkarni\inst{3}}

\institute{Astronomical Institute, Utrecht University, PO Box 80\,000,
           3508 TA Utrecht, The Netherlands\\
           \email{c.g.bassa@astro.uu.nl}
      \and Department of Astronomy and Astrophysics, University of
           Toronto, 60 Saint George Street, Toronto, ON M5S 3H8,
           Canada
      \and Palomar Observatory, California Institute of Technology
           105-24, Pasadena, CA 91125, USA
}

\offprints{C.G. Bassa}

\date{Received / Accepted}

\abstract{We present optical and near-infrared observations with Keck
  of the binary millisecond pulsar PSR~J0751+1807. We detect a faint,
  red object -- with $R=25.08\pm0.07$, $B-R=2.5\pm0.3$, and
  $R-I=0.90\pm0.10$ -- at the celestial position of the pulsar and
  argue that it is the white dwarf companion of the pulsar. The
  colours are the reddest among all known white dwarfs, and indicate a
  very low temperature, $T_\mathrm{eff}\approx4000$\,K. This implies that
  the white dwarf cannot have the relatively thick hydrogen envelope
  that is expected on evolutionary grounds. Our observations pose two
  puzzles. First, while the atmosphere was expected to be pure
  hydrogen, the colours are inconsistent with this
  composition. Second, given the low temperature, irradiation by the
  pulsar should be important, but we see no evidence for it. We
  discuss possible solutions to these puzzles.

}

\maketitle

\section{Introduction}\label{sec:intro}
Among the pulsars in binaries, the largest group, the low-mass binary
pulsars, has low-mass white-dwarf companions. Before the companions
became white dwarfs, their progenitors filled their Roche lobe and
mass was transferred to the neutron stars, thereby spinning them up and
decreasing their magnetic fields.  Considerations of the end of this
stage, where the white dwarf progenitor's envelope becomes too tenuous
to be supported further, allow one to make predictions for relations
between the orbital period and white dwarf mass, and orbital period
and eccentricity (for a review, e.g., \citealt{pk94,sta04}). Furthermore,
after the cessation of mass transfer, two clocks will start ticking at
the same time: the neutron star, now visible as a millisecond pulsar,
will spin down, while the secondary will contract to a white dwarf and
start to cool.  Consequently, the spin-down age of the pulsar should
equal the cooling age of the white dwarf.

From optical observations of white-dwarf companions to millisecond
pulsars one can estimate the white-dwarf cooling age and compare it
with the pulsar spin-down age. Initial attempts to do this
\citep{hp98a,hp98b,sdb00} revealed a dichotomy in the cooling
properties of white dwarfs in the sense that some white dwarf
companions to older pulsars have cooled less than those of younger
pulsars. In particular, the companions of PSR~J0437$-$4715
\citep{dbv93,sbb+01} and PSR~B1855+09 \citep{kbkk00,rt91} have
temperatures of about 4000--5000\,K, with characteristic pulsar ages
of 5\,Gyr. This is in contrast to the companion of PSR~J1012+5307
\citep{llfn95,kbk96,cgk98}, which has a higher temperature (8600\,K),
while it orbits an older pulsar (8.9\,Gyr).

A likely cause for this dichotomy is the difference in the thickness
of the envelope of hydrogen surrounding the helium core of the white
dwarf \citep{ashp96}. After the cessation of mass transfer, the white
dwarfs have relatively thick ($\sim\!10^{-2}$\,M$_\odot$) hydrogen
envelopes which are able to sustain residual hydrogen shell-burning,
keeping the white dwarf hot and thereby slowing the cooling
\citep{dsbh98}.  The shell burning, however, can become unstable and
lead to thermal flashes which can reduce the mass of the
envelope. White dwarfs with such reduced, relatively thin
($\la10^{-3}$\,M$_\odot$) hydrogen envelopes cannot burn hydrogen and,
as a result, cool faster. The transition between thick and thin
hydrogen envelopes was predicted to lie near 0.18--0.20\,M$_\odot$
(where heavier white dwarfs have thin envelopes;
\citealt{ashp96,seg00,asb01}).

Until recently, PSR~J1012+5307, with an orbital period
$P_\mathrm{b}=0.60$\,d, was the only system for which a thick hydrogen
envelope was required to match the two timescales. Given the relation
between the white dwarf mass and the orbital period
\citep{jrl87,rpj+95,ts99}, companions in similar or closer orbits
should have similar or lower mass, and thus have thick hydrogen
envelopes as well. This was confirmed by the recent discovery of two
new, nearby, binary millisecond pulsars with orbital periods similar
to that of PSR~J1012+5307; PSR~J1909$-$3744 (1.53\,d,
\citealt{jhb+05}) and PSR~J1738+0333 (0.354\,d, Jacoby et al., in
prep.; see \citealt{kbjj05} for preliminary results). For both, the
temperatures and characteristic ages are similar to those of PSR
J1012+5307, and thus one is led to the same need for a thick hydrogen
envelope. These discoveries, combined with the thin envelopes inferred
for PSR~J0034$-$0534 (1.59\,d) and binaries with longer periods,
suggest that the transition occurs at a mass that corresponds to an
orbital period just over 1.5\,d (\citealt{kbjj05}). All systems with
shorter orbital periods should have thick hydrogen envelopes.

The two known millisecond pulsars with white dwarf companions that
have shorter orbital periods than PSR~J1012+5307 but do not have
optical counterparts, are PSR~J0613$-$0200, with a 1.20\,d period, and
PSR~J0751+1807, which has the shortest orbital period of all binary
millisecond pulsars with $M_\mathrm{c}>0.1$\,M$_\odot$ companions,
0.26\,d \citep{lzc95}. The latter system is of particular interest
because the companion mass has been determined from pulse timing
($M_\mathrm{WD}=0.19\pm0.03$\,M$_\odot$ at 95\% confidence;
\citealt{nss+05}), so that one does not have to rely on the
theoretical period-mass relationship. Intriguingly, for PSR
J0751+1807, optical observations from \citet{lcf+96} set a limit to
the temperature of 9000\,K, which is only marginally consistent with
it having a thick hydrogen envelope. Based on this, \citet{esa01},
suggested the hydrogen envelope may have been partially lost due to
irradiation by the pulsar.

The faintness of the companion to PSR~J0751+1807 aroused our curiosity
and motivated us to obtain deep observations to test the theoretical
ideas discussed above. We describe our observations in
Sect.~\ref{sec:observations}, and use these to determine the
temperature, radius and cooling history in Sect.~\ref{sec:tandr}. In
Sect.~\ref{sec:irradiation}, we investigate irradiation by the pulsar,
finding a surprising lack of evidence for any heating. We discuss our
results in Sect.~\ref{sec:discussion}.

\section{Observations and data reduction}\label{sec:observations}
The PSR~J0751+1807 field was observed with the 10~meter Keck I and II
telescopes on Hawaii on five occasions. On December 11, 1996 the Low
Resolution Imaging Spectrometer (LRIS, \citealt{occ+95}) was used to
obtain $B$ and $R$-band images, while the Echellette Spectrograph and
Imager (ESI, \citealt{sbe+02}) was used on December 21, 2003 to obtain
deeper $B$ and $R$-band, as well as $I$-band images. The $R$-band
filter used that night was the non-standard ``Ellis $R$'' filter. The
observing conditions during the 1996 night were mediocre, with
0\farcs8--1\farcs1 seeing and some cirrus appearing at the end of the
night. The conditions were photometric during the 2003 night, and the
seeing was good, 0\farcs6--0\farcs8. The third and fourth visit were
with LRIS again, now at Keck I, on January 7 and 8, 2005. The red arm
of the detector was used to obtain $R$-band images. The seeing on the
first night in 2005 was rather bad, about 1\farcs5 and improved to
about 1\farcs0 on the second night. The conditions on these nights
were not photometric. Finally, a series of 36 dithered exposures, each
consisting of 5 co-added 10\,s integrations, were taken through the
$K_\mathrm{s}$ filter with the Near Infrared Camera (NIRC;
\citealt{ms94}) on January 26, 2005. The conditions were photometric
with 0\farcs6 seeing. Standard stars (\citealt{lan92,ste00}) were
observed in 1996 and 2003, while a 2MASS star \citep{csd+03} in the
vicinity of PSR~J0751+1807 was observed to calibrate the NIRC data. A
log of the observations is given in Table~\ref{tab:obs}.

The images were reduced using the Munich Image Data Analysis System
(MIDAS). The $BRI$ images were bias-subtracted and flat-fielded using
dome flats. The longer exposures in each filter were aligned using
integer pixel offsets, and co-added to create average images. The
near-infrared images were corrected for dark current using dark frames
with identical exposure times and number of co-adds as those used for
the science frames. Next, a flatfield frame was created by median
combining the science frames. After division by this flatfield, the
science frames were registered using integer pixel offsets and
averaged.

\begin{table}
  \begin{minipage}[t]{\columnwidth}
  \centering
  \caption[]{Observation log.}\label{tab:obs}
  \renewcommand{\footnoterule}{}
  \begin{tabular}{l@{\hspace{0.5cm}}
      c@{\hspace{0.5cm}}
      c@{\hspace{0.5cm}}
      c@{\hspace{0.5cm}}
      c@{\hspace{0.5cm}}
    }
    \hline
    Field & Time (UT) & Filter & $t_\mathrm{int}$ (s) & $\sec z$ \\
    \hline
    \multicolumn{5}{l}{December 11, 1996, LRIS} \\[0.2ex]
    SA\,95         & 08:23--08:25 & $R$ & $2+10$ & 1.07 \\
                   & 08:27--08:29 & $B$ & $2+10$ & 1.07 \\[0.1ex]
    SA\,95         & 09:28--09:31 & $B$ & $2+10$ & 1.07 \\
                   & 09:33--09:35 & $R$ & $2+10$ & 1.08 \\[0.1ex]
    PSR~J0751+1807 & 09:45        & $R$ & $10$ & 1.39 \\
                   & 09:47--09:59 & $R$ & $2\times300$ & 1.36 \\
                   & 10:01        & $R$ & $600$ & 1.31 \\
                   & 10:13        & $B$ & $600$ & 1.26 \\[0.8ex]
    \multicolumn{5}{l}{December 21, 2003, ESI} \\[0.2ex]
    PSR~J0751+1807 & 10:06--10:27 & $R$ & $3\times360$ & 1.14 \\
                   & 10:29--10:57 & $I$ & $6\times240$ & 1.08 \\
                   & 11:00--11:33 & $B$ & $3\times600$ & 1.04 \\[0.1ex]
    NGC\,2419      & 11:40 & $B$ & $10+30$ & 1.06 \\
                   & 11:44 & $R$ & $10+30$ & 1.06 \\
                   & 11:47 & $I$ & $10+30$ & 1.06 \\[0.8ex]      
    \multicolumn{5}{l}{January 7, 2005, LRIS} \\[0.2ex]
    PSR~J0751+1807 & 11:54--12:53 & $R$ & $5\times600$ & 1.05 \\[0.8ex]
    \multicolumn{5}{l}{January 8, 2005, LRIS} \\[0.2ex]
    PSR~J0751+1807 & 11:42--12:51 & $R$ & $6\times600$ & 1.05 \\[0.8ex]
    \multicolumn{5}{l}{January 26, 2005, NIRC} \\[0.2ex]
    PSR~J0751+1807 & 08:06--08:56 & $K_\mathrm{s}$ & $36\times50$ & 1.07 \\ 
    2MASS star\footnote{2MASS\,J07510621+1807253} & 08:59 & $K_\mathrm{s}$ & $0.4$ & 1.02\\
    \hline
  \end{tabular}
  \end{minipage}
\end{table}

\subsection{Astrometry}
For the astrometric calibration, we selected 14 stars from the second
version of the USNO CCD Astrograph catalogue (UCAC2; \citealt{zuz+04})
that overlapped with the 10\,s $R$-band LRIS image of December
1996. Of these, 11 were not saturated and appeared stellar and
unblended. The centroids of these objects were measured and corrected
for geometric distortion using the bi-cubic function determined by
J.~Cohen (1997, priv.\
comm.)\footnote{http://alamoana.keck.hawaii.edu/inst/lris/coordinates.html}. We
fitted for zero-point position, plate scale and position angle. The
inferred uncertainty in the single-star measurement of these 11 stars
is $0\farcs057$ and $0\farcs083$ in right ascension and declination,
respectively, and is consistent with expectations for the UCAC
measurements of approximately $0\farcs020$ for stars of 14th magnitude
and $0\farcs070$ for stars 2 magnitudes fainter.

This solution was transferred to the 10\,min $R$-band LRIS image using
91 stars that were present on both images and were stellar,
unsaturated and not blended. Again the zero-point position, plate
scale and position angle were left free in the fit and the final
residuals were $0\farcs016$ and $0\farcs019$ in right ascension and
declination. The UCAC is on the International Celestial Reference
System (ICRS) to $\la\!0\farcs01$, and hence the final systematic
uncertainty with which our coordinates are on the ICRS is dominated by
our first step, and is $\sim\!0\farcs03$ in each coordinate.

\begin{figure}
  \resizebox{\hsize}{!}{\includegraphics{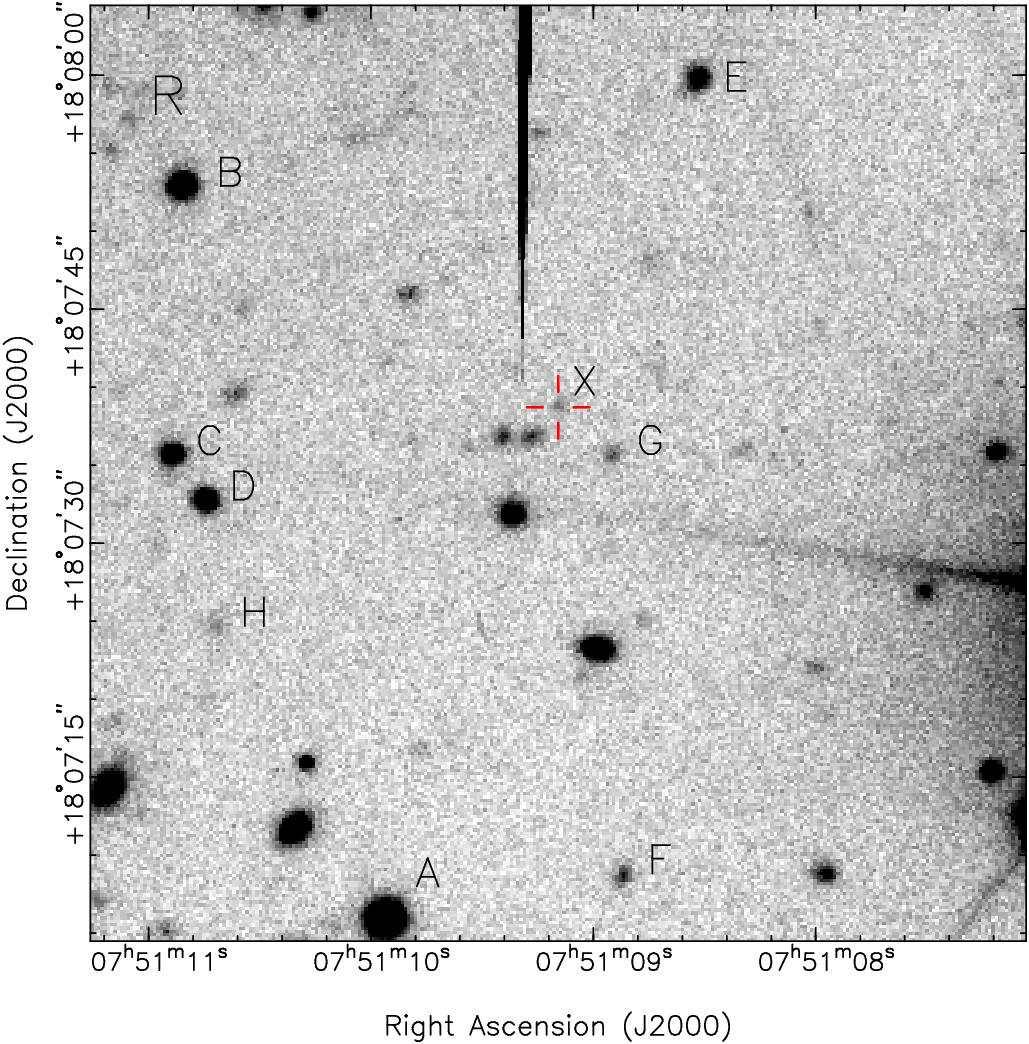}}
  \includegraphics[width=2.15cm]{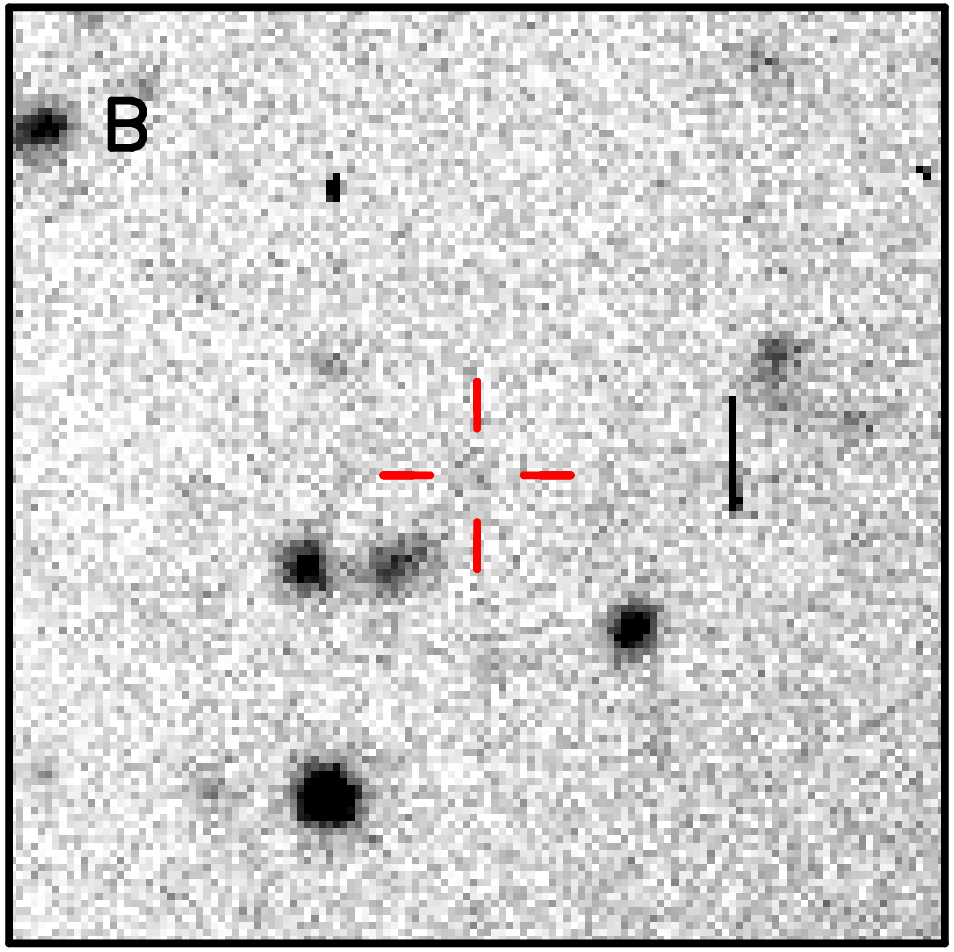}
  \includegraphics[width=2.15cm]{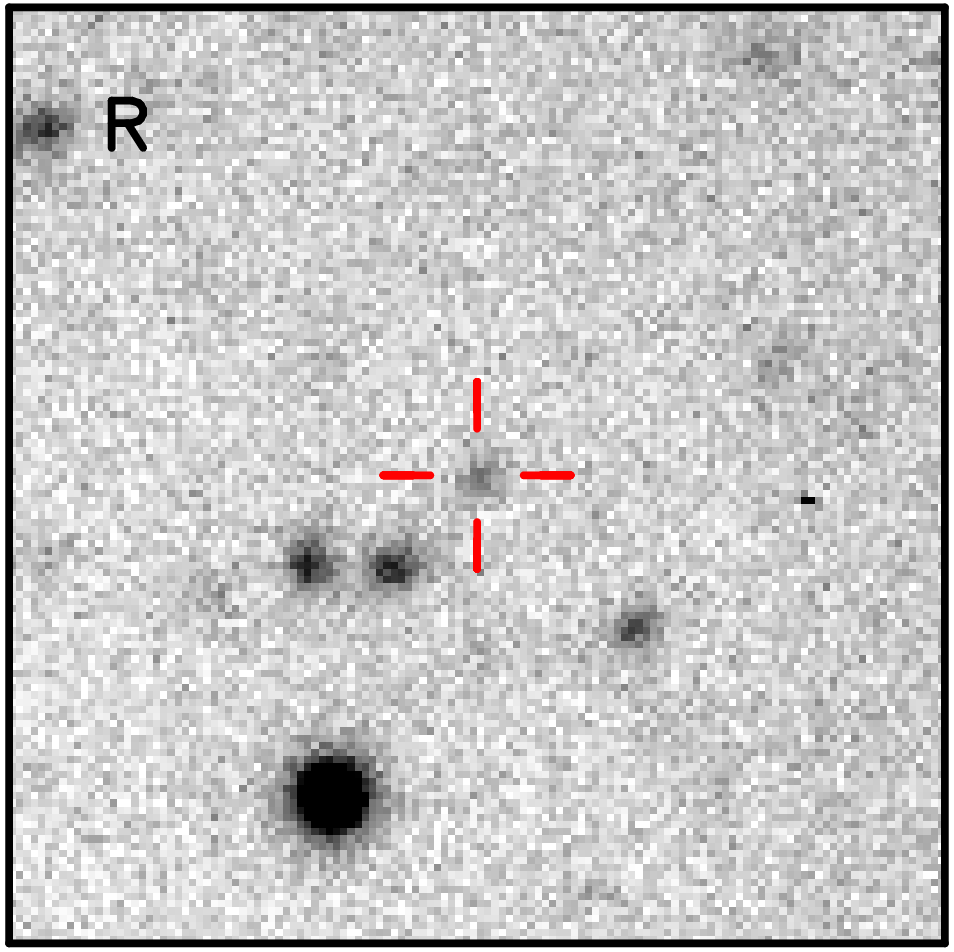}
  \includegraphics[width=2.15cm]{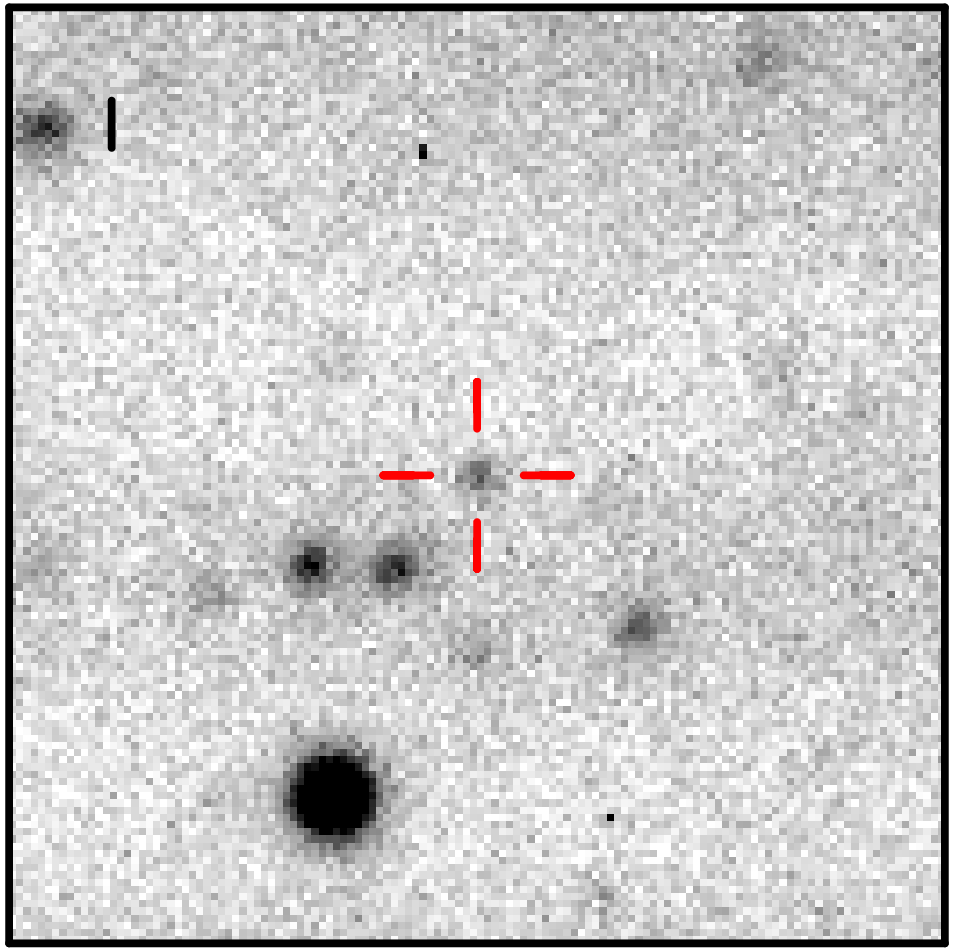}
  \includegraphics[width=2.15cm]{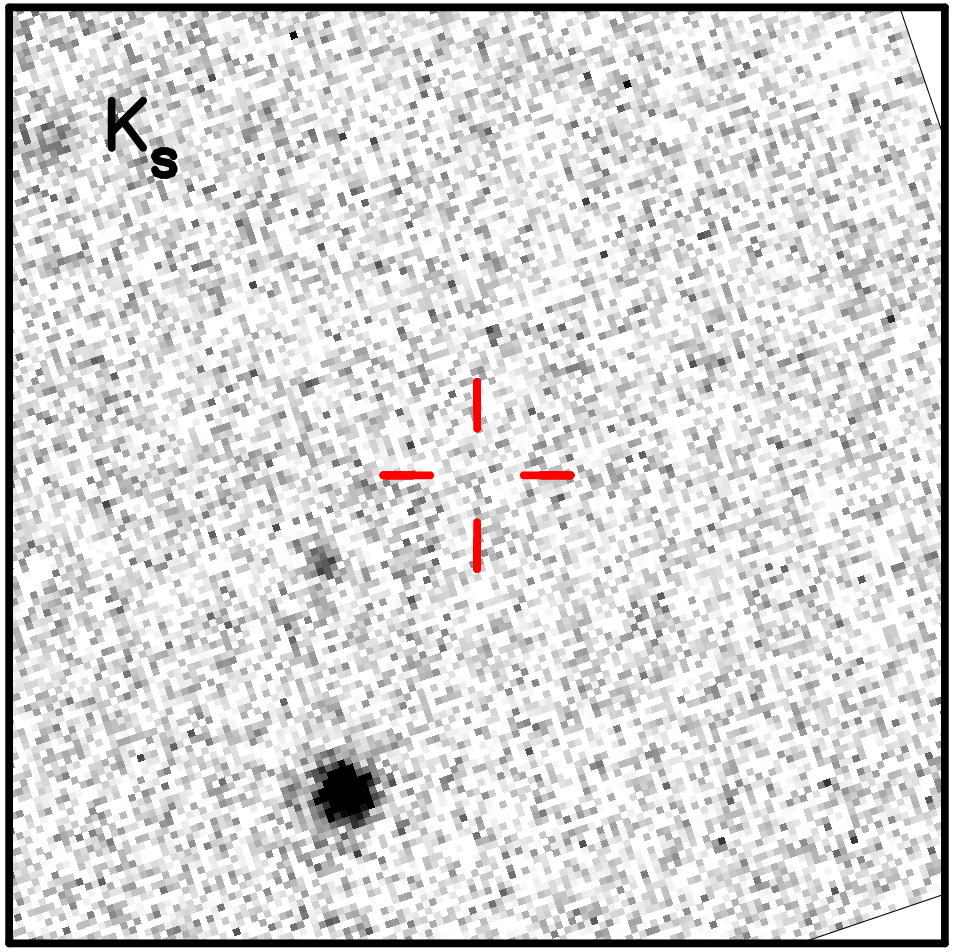}
  \caption{Images of the field of PSR~J0751+1807. The upper figure
  shows a $1\arcmin \times 1\arcmin$ subsection of the averaged
  $6\times10$\,m $R$-band image obtained with LRIS on January 8,
  2005. The bottom four figures show $20\arcsec \times 20\arcsec$
  subsections of the $B$, $R$ and $I$-band averages observed with ESI
  in 2003 and the $K_\mathrm{s}$-band image observed with NIRC in
  2005. The tick marks all have a length of $1\arcsec$ and are
  centered on the pulsar timing position.}
  \label{fig:fc}
\end{figure}

Our images, with the position of PSR~J0751+1807 \citep{nss+05}
indicated, are shown in Figure~\ref{fig:fc}. On the 10\,min LRIS
$R$-band images from 1996 and 2005, we find a faint object, hereafter
star X, at the position of the pulsar. It is also, though marginally,
present in the two 5\,min $R$-band images from 1996, but not detected
in the 10\,min $B$-band LRIS image of that observing run. Star X is
clearly present in the 2003 ESI $R$ and $I$-band images, and
marginally in the $B$-band image. It is not detected in the
near-infrared observations (Fig.~\ref{fig:fc}).

Positions for star X and other objects inferred using the astrometry
of the 10\,min LRIS $R$-band image are listed in Table~\ref{tab:phot}.
The pulsar position at the time of the 1996 LRIS observation, using
the \citet{nss+05} position and proper motion, is
$\alpha_\mathrm{J2000}=07^\mathrm{h}51^\mathrm{m}09\fs1574(1)$,
$\delta_\mathrm{J2000}=+18\degr07\arcmin38\farcs624(10)$.  We find
that star X is offset from the pulsar position by
$-0\farcs01\pm0\farcs06$ in right ascension and
$0\farcs04\pm0\farcs06$ in declination, well within the $1\sigma$
uncertainties (including those on the pulsar position). Given the low
density of about 47 stars per square arcminute and the excellent
astrometry, the probability of a chance coincidence in the 95\%
confidence error circle, which has a radius of $0\farcs24$, is only
0.1--0.2\%. Since, as we will see, it is hard to envisage how the
companion could be fainter than the object detected, we are confident
that star X is the companion of PSR~J0751+1807.

\subsection{Photometry}\label{sec:photometry}
The DAOPHOT II package (\citealt{ste87}), running inside MIDAS, was
used for the photometry on the averaged 
images. We followed the recommendations of
\citet{ste87}: instrumental magnitudes were obtained through point
spread function (PSF) fitting and aperture photometry on brighter
stars was used to determine aperture corrections. 

For the calibration of the optical images, instrumental magnitudes of
the standard stars, determined using aperture photometry, were
compared against the values of \citet{ste00}. We used the standard
Keck extinction coefficients of 0.17, 0.11 and 0.07\,mag per airmass
for $B$, $R$ and $I$, respectively. Colour terms were not required for
the LRIS $B$ and $R$ bands, but were significant for the ESI bands:
$0.107 (B-R)$ for $B$, $0.083 (B-R)$ for $R$, and $-0.004 (R-I)$ for
$I$, i.e., the ESI $B$, $R$ are redder than the standard bands, while
ESI $I$ is slightly bluer. The root-mean-square residuals of the ESI
calibrations are about 0.05\,mag in $B$, and 0.03\,mag in $R$ and $I$,
while those of the LRIS calibration are 0.08\,mag in $B$ and 0.05\,mag
in $R$; we adopt these as the uncertainty in the zero-points. The
near-infrared observations were calibrated through aperture photometry
with 1\farcs5 (10\,pix) apertures using the 2MASS star, fitting for a
zero-point only, as the difference in airmass between the science and
calibration images is small. We adopt an uncertainty in the
$K\mathrm{s}$ zero-point of 0.1\,mag.

Calibrated ESI magnitudes for star X and selected other stars in the
field are listed in Table~\ref{tab:phot}. Star X is barely above the
detection limit of the ESI $B$-band observations, hence the large
error. It is not detected in the LRIS $B$-band and the NIRC
$K_\mathrm{s}$-band observations, and, scaling from the magnitude of a
star with a signal-to-noise ratio of about 10 and 6, we estimate the
3$\sigma$ detection limits at $B=26.8$ and $K_\mathrm{s}=21.3$,
respectively.  The former is consistent with the ESI detection. None
of the stars in Table~\ref{tab:phot} are covered by the small
$38\arcsec\times38\arcsec$ field-of-view of NIRC, hence we do not have
near-infrared magnitudes for these.

The 1996 LRIS $R$-band magnitude is $25.13\pm0.11$, which is
consistent with the ESI measurement. Since the conditions during the
1996 LRIS observations may not have been photometric, however, this
may be a coincidence. To check for variability, we tied the
instrumental LRIS $R$ band magnitudes directly to the ESI $R$ and $I$
ones, using 38 stars that both images had in common and that had
magnitude uncertainties below 0.1 mag. As expected given the
non-standard ``Ellis $R$'' filter on ESI, we required a large colour
term, $-0.302(R_\mathrm{inst}-I_\mathrm{inst})$, but with this the fit
was adequate, with root-mean-square residuals of 0.14 mag. Compared to
the fit, the ESI minus LRIS difference in $R$-band magnitude is
insignificant, $-0.03\pm0.13$\,mag. Similarly, comparing instrumental
$R$-band magnitudes from 2005 January 7 with those taken 2005 January
8 and 1996 December 11, fitting for an offset only, results in
magnitude differences of $0.03\pm0.07$ and $-0.16\pm0.12$\,mag,
respectively.  Thus, no large variations in brightness are seen; we will
see in Sect.~\ref{sec:irradiation} that this is somewhat surprising.

\begin{table}
  \caption[]{LRIS Astrometry and ESI photometry of the companion of
    PSR~J0751+1807 and stars in the field. The nomenclature of the
    stars is according to Fig.~\ref{fig:fc}. The uncertainties
    listed in parentheses are instrumental, i.e., they do not include
    the zero-point uncertainty in the astrometric tie (about
    $0\farcs03$ in each coordinate) or of photometric calibration
    (0.05 mag in $B$ and 0.03 mag in both $R$ and $I$).}
  \label{tab:phot}
  \begin{tabular}
    {l@{\hspace{0.15cm}}
     l@{\hspace{0.15cm}}
     l@{\hspace{0.15cm}}
     l@{\hspace{0.15cm}}
     l@{\hspace{0.15cm}}
     l@{\hspace{0.15cm}}
    }
    \hline
    \hline
    ID & \multicolumn{1}{c}{$\alpha_\mathrm{2000}$} & 
    \multicolumn{1}{c}{$\delta_\mathrm{2000}$} & 
    \multicolumn{1}{c}{$B$\phantom{0}} & 
    \multicolumn{1}{c}{$R$\phantom{0}} &
    \multicolumn{1}{c}{$I$\phantom{0}} \\
     & $\phantom{00}^\mathrm{h}\phantom{00}^\mathrm{m}\phantom{00}^\mathrm{s}$
     &$\phantom{00}\degr\phantom{00}\arcmin\phantom{00}\arcsec$ & & & \\
    \hline
    X &  07 51 09.158(4) &  18 07 38.66(6)  & 27.56(25) & 25.08(7) & 24.18(7) \\[0.2em]
    A &  07 51 09.933(1) &  18 07 05.97(1)  & 21.73(1)  & 19.30(1) & 18.31(1) \\
    B &  07 51 10.844(1) &  18 07 52.91(1)  & 22.80(1)  & 21.03(1) & 20.32(1) \\
    C &  07 51 10.891(1) &  18 07 35.69(1)  & 24.30(2)  & 21.81(1) & 20.63(1) \\
    D &  07 51 10.739(1) &  18 07 32.79(1)  & 24.28(6)  & 22.50(5) & 21.99(6) \\
    E &  07 51 08.519(1) &  18 07 59.89(2)  & 24.56(7)  & 22.87(5) & 22.38(8) \\
    F &  07 51 08.859(2) &  18 07 08.83(3)  & 24.94(4)  & 24.00(5) & 23.29(4) \\
    G &  07 51 08.908(4) &  18 07 35.71(5)  & 25.65(8)  & 24.51(5) & 23.85(6) \\
    H &  07 51 10.691(3) &  18 07 24.69(6)  & 25.69(7)  & 24.94(9) & 24.34(8) \\
    \hline
  \end{tabular}
\end{table}

\section{Temperature, radius, and cooling history}\label{sec:tandr}

We use our observations of star X, the companion of PSR~J0751+1807, to
constrain its temperature, radius, and atmospheric constituents, and
discuss our result that the white dwarf does not have the expected
thick hydrogen envelope.

\subsection{Colours, temperature, and atmospheric composition}
We first use the colours of star X to constrain its temperature. The
red colours are largely intrinsic, as the maximum reddening towards
PSR~J0751+1807 ($l=202.73$, $b=21.09$) is small, $E_{B-V}=0.05\pm0.01$
\citep{sfd98}. This value is consistent with the low value found for
the interstellar absorption
$N_\mathrm{H}\sim4\times10^{20}$\,cm$^{-2}$, as estimated from {\it
ROSAT} X-ray observations of PSR~J0751+1807 by \citet{btl+96}. For
comparison, the relation by \citet{ps95} predicts an
$N_\mathrm{H}\approx3\times10^{20}$\,cm$^{-2}$ for the above
reddening. Given the distance of $\sim\!0.6\,$kpc \citep{nss+05}, we
expect most of the reddening to be in the foreground to the
pulsar. Hence, the dereddened colours are $(B-R)_0=2.40\pm0.27$ and
$(R-I)_0=0.86\pm0.10$.

\begin{figure}
  \centerline{\resizebox{\hsize}{!}{\includegraphics{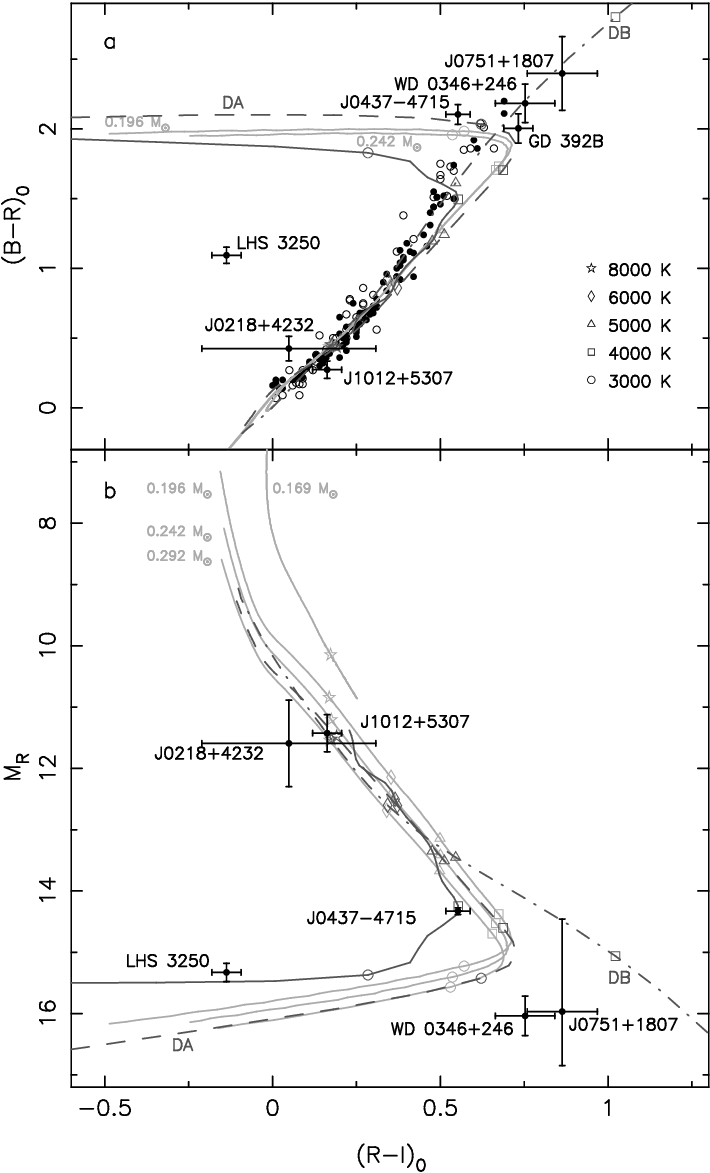}}}
  \caption{{\bf a} Colour-colour and {\bf b} colour-magnitude diagram
    for the companion of PSR~J0751+1807, other millisecond pulsar
    companions, field white dwarfs, and model predictions. Shown with
    error bars are PSR~J0437$-$4715 \citep{dbv93}, PSR~J1012+5307
    \citep{llfn95}, PSR~J0218+4232 \citep{bkk03} and PSR~J0751+1807
    (this work), as well as the ultra-cool field white dwarfs LHS 3250
    \citep{hdv+99}, WD~0346+246 \citep{osh+01}, and GD 392B
    \citep{far04}. In the colour-colour diagram, also the full sample
    of field white dwarfs of \citet{blr01} is shown, with filled and
    open circles indicating white dwarfs with and without H$\alpha$ in
    their spectrum, respectively. For the colour-magnitude diagram, we
    used parallax distances where available and distances inferred
    from the pulsar dispersion measure otherwise; we omitted the
    \citeauthor{blr01} sample because of the large scatter (even
    though it likely is largely intrinsic). In both panels, the
    continuous light grey lines depict helium-core white dwarf cooling
    models by \citet{sarb01}, with masses as indicated. The continuous
    dark-grey track is for a 0.2\,M$_\odot$ model by B.~Hansen (priv.\
    comm.; see description in \citealt{hp98a}). The dark-grey dashed
    and dashed-dotted lines are updates of the white dwarf models by
    \citet{bwb95}, for DA (hydrogen rich) and DB (helium rich)
    composition, respectively, both with $\log g=7$. Temperatures for
    all models are indicated by different symbols along the track.}
  \label{fig:tcd}
  \label{fig:cmd}
\end{figure}

In Fig.~\ref{fig:tcd}a, we compare the intrinsic colours of star X with
those of other white-dwarf companions of millisecond pulsars, other
white dwarfs, and models. We find that the colours of star X are the
reddest for any known millisecond pulsar companion or white dwarf.
The pulsar companion that comes closest is that of PSR~J0437$-$4715
($B-R=2.12\pm0.06$, $R-I=0.56\pm0.02$ [\citealt{dbv93}] and negligible
extinction\footnote{As inferred from the dust maps of \cite{sfd98};
\cite{dbv93} estimate $E_{B-V}=0.07$ from the work of \cite{knu79}.});
the most similar white dwarf is WD~0346+246 ($B-R=2.2\pm0.1$,
$R-I=0.76\pm0.08$, \citealt{osh+01}). Thus, star X is likely as cool
or even cooler than the $T_{\rm eff}\simeq3700\,$K inferred for those
two sources (PSR~J0437$-$4715: \citealt{dbv93}; Hansen 2002, priv.\
comm.; WD~0346+246: \citealt{osh+01,ber01}).

Also shown in Fig.~\ref{fig:tcd}a are colours expected from model
atmospheres of \citet{sarb01} and of Hansen (2004, priv.\ comm.),
which are specifically tailored to the low-mass, helium-core
companions of millisecond pulsars, as well as those for updated
low-gravity ($\log g=7$), pure hydrogen atmosphere models\footnote{For
updated versions of the \citet{bwb95} models, see
http://www.astro.umontreal.ca/\~{}bergeron/CoolingModels/} of
\citet{bwb95}. One sees that the colours of the companion of
PSR~J0437$-$4715, as well as those of the hotter companions of
PSR~J1012+5307 and J0218+4232, are consistent with these models. For
star X, however, the colours are not consistent, as the models never
venture redwards of $R-I\approx0.7$ and $B-R\approx2.0$.

The change in direction of the tracks is seen in all models for
hydrogen-rich, metal-free atmospheres; it reflects a change in the
dominant source of opacity, from bound-free absorption of H$^-$ at
higher temperatures to collision-induced absorption of H$_2$ at lower
ones (\citealt{lcs91,sbl+94,han98}). The latter process is highly
non-grey, and leads to absorption predominantly longward of the
$R$-band. As a result, the $R-I$ colour becomes bluer with decreasing
temperatures, while $B-R$ remains roughly constant.

Could star X have a different composition? Due to the high gravity of
white dwarfs, metals settle out of the atmosphere. However, some white
dwarfs have atmospheres dominated not by hydrogen, but by helium. For
the latter, the opacity sources are all fairly grey, and hence the
colours continue to redden with decreasing temperatures.  Indeed, the
colours of star X are consistent with the predictions of the updated
$\log g=7$ pure helium models after \citet{bwb95} at
$T_\mathrm{eff}\simeq4200$\,K (Fig.~\ref{fig:tcd}a).

From an evolutionary perspective, however, a pure helium atmosphere is
not expected. Low-mass white dwarfs such as the companions to
millisecond pulsars are all formed from low-mass stars whose evolution
was truncated by mass transfer well before helium ignition (for recent
models, see \citealt{ts99,ndm04}). As a result, they should have
helium cores surrounded by relatively thick, 0.01 to 1\% of the mass,
hydrogen envelopes (\citealt{dsbh98,asb01}). Indeed, among the
low-mass white-dwarf companions to pulsars \citep{kbjj05} as well as
among low-mass white dwarfs in general \citep{blr01}, only
hydrogen-dominated atmospheres have been observed.

In principle, at low temperatures, the hydrogen envelope might become
mixed in with the helium core. Even if fully mixed, however, the
remaining amounts of hydrogen would strongly influence the spectrum.
Indeed, the effects of collision-induced absorption {\em increase}
with increasing helium abundance up to $N({\rm He})/N({\rm
H})\simeq10^5$ (\citealt{bl02}).

From Fig.~\ref{fig:tcd}a, it is clear that the predictions for
hydrogen-dominated atmospheres are also a somewhat poor match to the
colours of the cooler normal white dwarfs with hydrogen in their
atmospheres (as inferred from absorption at H$\alpha$,
\citealt{blr01}; filled circles in the figure). For most, this appears
to be due to missing blue opacity in the models (see \citealt{blr01}
for a detailed study); the visual through infrared fluxes are
reproduced well by the models, and show unambiguously that
collision-induced absorption by H$_2$ is important. Indeed, the
absorption is evident in the optical colours of some objects, in
particular LHS~3250 (shown in Fig.~\ref{fig:tcd}) and
SDSS~J133739.40+000142.8 (\citealt{bl02} and references therein).

For our purposes, however, the case of the ultra-cool white dwarf
WD~0346+246 is most relevant. For this source, the colours cannot be
reproduced with either pure hydrogen or helium, but require a mixed
atmosphere, dominated by helium (with fractional hydrogen abundances
ranging from $10^{-9}$ to $10^{-1}$, depending on assumptions about
the contribution of other opacity sources; \citealt{osh+01,ber01},
though recent work puts these abundances in to doubt, P.\ Bergeron
2005, priv.\ comm.).  For all cases, the temperature is around
3700\,K. The similarity in the colours of WD~0346+246 and star~X would
suggest that star~X has a similar, maybe slightly lower, temperature.

From the above, we find that we cannot determine the temperature of
the companion of PSR~J0751+1807 with certainty, since we do not know
its composition. Most likely, however, it is somewhere between the
temperature inferred for WD~0346+246 and that indicated by the (pure
helium) models, i.e.\ in the range of, say 3500--4300\,K.

A more stringent test could be provided by the near-infrared
observations, as the $R-K$ colour (which is similar to
$R-K_\mathrm{s}$) differs for different predictions. At a temperature
of 4000\,K the $\log g=7$ \citet{bwb95} models predict $R-K$ colours
of 2.7 and 1.6 for pure helium and pure hydrogen atmospheres,
respectively. For the same temperature, $R-K=1.6$ is predicted by the
0.196\,M$_\odot$ model by \citet{sarb01}. Finally, for WD~0346+246,
with presumably a mixed hydrogen/helium atmosphere, \citet{osh+01}
observed $R-K=-0.7$. Unfortunately, our near-infrared observations
only limit the colour to $R-K<3.8$, which does not constrain any of
these predictions.

\subsection{Brightness, distance and radius}
So far, we have only discussed the colours and temperature. We now
turn to the absolute magnitude and radius. In Fig.~\ref{fig:cmd}b, we
show $M_R$ as a function of $R-I$. For star~X, we computed the
absolute $R$-band magnitude $M_R$ using the parallax of
$\pi=1.6\pm0.8$\,mas as measured through radio timing \citep{nss+05}.
The resulting distance of $0.6^{+0.6}_{-0.2}$\,kpc is consistent with
that estimated from the dispersion measure which predicts
$1.1\pm0.2$\,kpc, using a dispersion measure of
$30.2489\pm0.003$\,pc\,cm$^{-3}$ \citep{nss+05} and the recent model
of the Galactic electron distribution of \citet{cl02}. Correcting for
the reddening, this implies $M_R=15.97^{+0.88}_{-1.51}$.

Given the similarities in the above absolute magnitude of star~X and
that of WD~0346+246 ($M_R=16.1\pm0.3$; \citealt{hsh+99,osh+01}), and
assuming similar temperature, one finds that the radius of star~X
should be comparable to the $R=0.010$\,R$_\odot$ for WD~0346+246
\citep{ber01}. However, the large uncertainty in the parallax of
PSR~J0751+1807 allows radii between 0.007--0.021\,R$_\odot$. For the
white-dwarf mass of $\sim\!0.19$\,M$_\odot$ inferred from pulse timing
\citep{nss+05}, this is consistent the $\sim\!0.022$\,R$_\odot$
expected from the 0.196\,M$_\odot$ model by \citep{sarb01}.

As can be seen in Fig.~\ref{fig:cmd}, the absolute magnitude is also
consistent with the predicted values from the $\log g=7$ pure helium
model by \citet{bwb95}. At a temperature of $T_\mathrm{eff}=4250$\,K,
this model has a radius of 0.020\,R$_\odot$ and a mass of
0.15\,M$_\odot$, somewhat smaller than the observed
0.19\,M$_\odot$. To correct for the small difference in mass, we
computed white dwarf radii for the observed temperature and mass of
the companion and used these to scale the absolute magnitudes of the
pure helium track in Fig.~\ref{fig:cmd}. At 0.19\,M$_\odot$ and
$T_\mathrm{eff}=4000$\,K, the \citet{pab00} helium core white dwarf
mass-radius relation predicts 0.021\,R$_\odot$. This is very similar
to the radius predicted by the \citet{bwb95} $\log g=7$ pure helium
models, and as such, the absolute magnitudes are comparable.  We
conclude that, with in the large uncertainties on the parallax
distance, the absolute magnitude and radius that we derive for the
companion of PSR~J0751+1807 are consistent with the predictions for a
pure helium atmosphere.

We note that of the models presented in Fig.~\ref{fig:cmd}, those of
\citet{bwb95} have been extensively tested to explain the population
of nearby white dwarfs \citep{blr01,ber01,bl02} and use a very
detailed description of the white dwarf atmosphere combined with the
latest opacities (P.\,Bergeron 2005, priv.\ comm.). This is not the
case for the models of Serenelli et al.\ and Hansen, and thus we
should be careful in using their models quantitatively. Indeed, as can
be seen from Fig.~\ref{fig:cmd}, their models do not reproduce the
observations of cool white dwarfs well. For instance, for the
companion of PSR~J0437$-$4715, which has a well-determined mass of
$0.236\pm0.017$\,M$_\odot$ and distance of $139\pm3\,$pc
\citep{sbb+01}, the models of \cite{sarb01}, while consistent with the
observed $B-R$ and $R-I$ well, do not reproduce $R-I$ and $M_R$
simultaneously. In contrast, the $0.2$\,M$_\odot$ model of Hansen
(2004, priv.\ comm.)  does pass through the $R-I$, $M_R$ point, but
cannot reproduce both colours.  It may be that both problems reflect
uncertainties in the model atmospheres used by Hansen and
\citet{sarb01}. It would be worthwhile to couple the evolutionary
models of these authors with the updated, very detailed atmospheric
model of \citet{bwb95}.

\subsection{Cooling history and nature of the envelope}\label{sec:cooling}
Despite the uncertainty in the models and in the composition of the
atmosphere, our observations show that the companion of PSR~J0751+1807
has cooled much more than expected if the amount of hydrogen was thick
enough for significant residual nuclear burning
(Sect.~\ref{sec:intro}). Indeed, the temperature is as expected if no
residual hydrogen burning occurred. For instance, at the
characteristic age of the pulsar, $\tau=7.1$\,Gyr \citep{nss+05}, the
0.196\,M$_\odot$ of \citet{sarb01}, which has a thin envelope,
predicts a temperature of about 3200\,K, which is roughly consistent
with what is observed. With a pure helium atmosphere, a slightly
colder temperature, of $\sim\!2500$\,K, is expected, though this is a
less secure estimate due to uncertainties in the opacities
\citep{hp98a}

\begin{figure}
  \resizebox{\hsize}{!}{\includegraphics{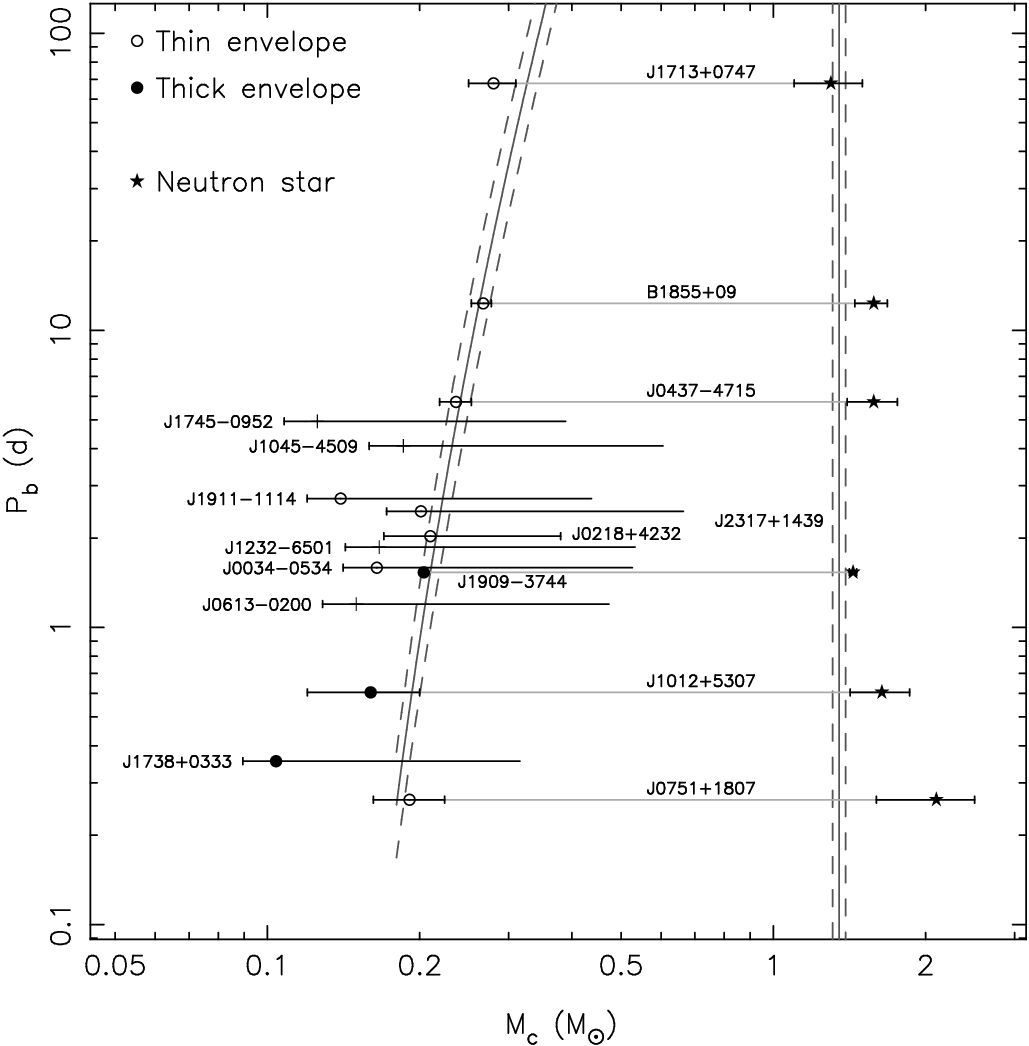}}
  \caption{The orbital period as a function of companion mass for a
  selection of low-mass binary pulsars outside globular clusters
  (either with $P_\mathrm{b}<5$\,d and
  $M_\mathrm{WD,min}>0.1$\,M$_\odot$, or with $P_\mathrm{b}<100$\,d
  and a secure companion mass determination). The data is compiled
  from the ATNF Pulsar Catalogue \citep{mhth05}, \citet{sta04} and
  \citet{kbjj05} and references therein. Companion masses are either
  determined (double error bars, 95\% confidence) or based on the
  assumption of a 1.35\,M$_\odot$ neutron star and an inclination of
  $i=90\degr$ (left error bar, minimum mass), 60\degr\ (central
  symbol, median mass) or 18\degr\ (right side, 5\% probability that
  $i$ is lower than this value and that the companion is
  heavier). Different central symbols indicate companions for which a
  thick or thin hydrogen envelope is inferred from optical
  measurements.  The systems for which the neutron star mass is
  measured are indicated and connected by a gray line with their
  companions. The vertical grey lines indicate the
  $1.35\pm0.04$\,M$_\odot$ neutron star mass range determined by
  \citet{tc99}, while the curved grey lines represent the theoretical
  relation by \citet{ts99} between the white dwarf mass and the
  orbital period.}
  \label{fig:pbm2}
\end{figure}

The presence of a thin (or no) hydrogen envelope is not expected,
however, since thick envelopes are inferred for other optically
identified companions in short-period systems (see
Sect.~\ref{sec:intro}). What could be wrong with this expectation? It
was based on two theoretical ideas: (i) that below a certain critical
mass, no shell flashes occur and hydrogen layers will be thick; and
(ii) that the companion mass monotonously increases with increasing
orbital period. These assumptions appeared to be confirmed by the
available data: for PSR~J0751+1807, with a period of 0.26\,d, the
companion mass of 0.16--0.21\,M$_\odot$ (95\% conf.; \citealt{nss+05})
is similar to what is found for two other short-period systems with
companions for which thick hydrogen envelopes are inferred, and less
than the masses for longer period systems with thin-envelope
companions. Specifically, PSR~J1012+5307 (0.60\,d,
0.12--0.20\,M$_\odot$) and PSR~J1909$-$3744 (1.53\,d,
0.19--0.22\,M$_\odot$) have thick envelopes while PSR~J0437$-$4715
(5.74\,d, 0.20--0.27\,M$_\odot$) and PSR~B1855+09 (12.33\,d,
0.24--0.29\,M$_\odot$) have thin envelopes (see Fig.~\ref{fig:pbm2}
and \citealt{kbjj05} and reference therein).  Thus, while the
uncertainties do not exclude that the companion of PSR~J0751+1807 is
so massive that it its envelope was diminished by shell flashes, the
existing data make it unlikely.

Two explanations for a thin envelope remain. First, there may be
differences in metallicity among the progenitors of pulsar companions.
\citet{sarb02} studied the evolution of low-mass pulsar companions
with sub-solar metallicity and found that, since the thermonuclear
flashes are induced by the reactions of the CNO-cycle, the threshold
mass between thin and thick hydrogen envelopes increases with
decreasing metallicity of the white dwarf progenitor. Thus, it may be
that the companion of PSR~J0751+1807 had a sufficiently higher
metallicity that it was above the threshold for shell flashes, while
companions in other short-period systems had lower metallicity and
hence were below the threshold, despite having higher masses.

The next possibility is that the white dwarf was indeed formed with a
thick envelope, which was subsequently removed by an action other than
shell flashes. Based on the upper limit on the temperature of
\citet{lcf+96}, \citet{esa01} already argued that the pulsar companion
could not have the thick hydrogen envelope, and they proposed a
scenario where part of the envelope was removed by pulsar
irradiation. \citeauthor{esa01} found that irradiation driven
mass-loss could remove as much as 0.01\,M$_\odot$ from the thick
hydrogen envelope (mostly while the companion is contracting following
the cessation of mass transfer).

A possible problem with the above suggestions, is that none predict
the removal of the entire hydrogen envelope, while the observed
colours seem most consistent with a pure helium or at least
helium-dominated atmosphere.

\section{Irradiation by the pulsar?}\label{sec:irradiation}
Above, we have treated the companion as if it were an isolated
object rather than member of a binary system. Might the presence of a
relatively energetic pulsar influence our observations?
 
The observed pulsar period and period derivative imply a spin-down
luminosity $L_\mathrm{SD}=(2 \pi)^2 I
\dot{P}/P^3=7.5\times10^{33}\,I_{45}$\,ergs\,s$^{-1}$ \citep{lzc95,nss+05},
where $I=10^{45}\,I_{45}{\rm\,g\,cm^2}$ is the pulsar moment of
inertia. For a $2.1$\,M$_\odot$ pulsar and a $0.19$\,M$_\odot$
companion, the orbital separation is $a=2.3$\,R$_\odot$, and,
consequently, the irradiative flux of the pulsar wind incident on the
companion is $f_{\rm
irr}=2.1\times10^{10}\,I_{45}$\,erg\,s$^{-1}$\,cm$^{-2}$. This is about
twice the flux of the companion itself, $f_{\rm th}=\sigma
T_\mathrm{eff}^4=1.06\times10^{10}$\,erg\,s$^{-1}$\,cm$^{-2}$ for
$T_\mathrm{eff}=3700$\,K. Therefore, the presence of the pulsar and
its irradiation may be important.

Given the irradiation, one would expect the side of the companion
facing the pulsar to be brighter than the side facing away from it.
Thus, from Earth, the companion should appear faintest at phase 0.25
and brightest at phase 0.75 (using the convention that at phase 0, the
pulsar is at the ascending node). This is indeed seen in other pulsar
binaries, with the black widow pulsar PSR~B1957+20 perhaps the most
spectacular example \citep{vpasc+88,fgld88}.

For star X, assuming a fraction $\eta$ of the incident flux is
absorbed and reradiated as optical flux, the flux from the bright side
of the companion should be a factor $1+\frac{2}{3}\eta f_{\rm
irr}/f_{\rm th}$ brighter (here, the factor $\frac{2}{3}$ reflects
projection effects). Observationally, the inferred values of $\eta$
range from 0.1 to 0.6 (\citealt{ok03}, and references therein), and
thus one expects a maximum change in bolometric flux by a factor 1.13
to 1.8. For the $R$-band flux, the range is 1.2 to 2.2 (assuming it
scales like a black-body spectrum, $\propto\!T^6$ around 3700\,K). We
confirmed this using a detailed light-curve synthesis model (described
briefly in \citealt{sklk99}).

For star~X, no effect is seen. Using the PSR~J0751+1807 ephemeris from
\citet{nss+05}, we find that during the ESI $R$-band observations the
orbital phase ranged from 0.22 to 0.25, while the 1996 LRIS $R$-band
images were taken at phases 0.86--0.90, and the 2005 LRIS images at
phases 0.01--0.14 on January 7, and 0.77--0.93 on January 8. Thus,
these observations span the orbital phases necessary to test for any
modulation in brightness. Indeed, using the inclination inferred from
timing, $i=66^{+4}_{-7}\,$deg \citep{nss+05}, we find that during the
ESI observations only 4 to 5\% of the irradiated part of the companion
surface was in view, while during the 1996 LRIS observations is was
78\% to 85\%. As a consequence, we expect to see nearly the maximum
change in brightness. Nevertheless, in Sect.~\ref{sec:photometry}, we
found no significant variation, $R_{\rm LRIS}-R_{\rm
ESI}=0.03\pm0.13$; thus, to $\sim\!99$\% confidence, the variation is
smaller than 0.3\,mag, which implies $\eta<0.15$.

The lack of observed modulation could be taken to indicate that the
irradiation is not very effective, e.g., because the albedo is large
(i.e., $\eta$ is small), the pulsar emission is non-isotropic, or the
spin-down luminosity is overestimated. We believe these options are
not very likely (for a discussion in a slightly different context, see
\citealt{ok03}), which leads us to consider the only alternative, that
one of the assumptions underlying the above estimates is wrong.

In particular, we assumed implicitly that irradiated flux is
reprocessed and re-emitted instantaneously, i.e., transfer of flux
inside and around the companion are assumed to have negligible effect.
For the companions of black-widow pulsars, this is reasonable, since
for these relatively large objects, tides will have ensured
synchronous rotation. Any flux transfer would thus have to be due to
winds and/or convection, which plausibly happens on a timescale long
compared to the thermal time of the layer in which the pulsar flux is
reprocessed.

The companion of PSR~J0751+1807, however, is well within its
Roche-lobe, and tidal dissipation should be negligible. We can
estimate its current rotation period from its prior evolution,
following the reasoning used by \cite{kk95} for the companion of PSR
B0655+64. Briefly, during mass transfer, the companion filled its
Roche-lobe and tides ensured the system was synchronised and
circularised. Once mass transfer ceased and the companion started to
contract to a white dwarf, however, the tides became inefficient, and
the rotational evolution of the companion was determined by
conservation of angular momentum.

For our estimates, we split the total moment of inertia of the
progenitor into two parts, one from the core, $I_\mathrm{core}=k_{\rm
core}^2M_\mathrm{core} R_\mathrm{core}^2$ and one from the envelope,
$I_\mathrm{env}=k_{\rm env}^2M_\mathrm{env} R_\mathrm{L}^2$; here $k$
is the radius of gyration and $R_{\rm L}$ is the radius of the Roche
lobe. After contraction of the envelope, one is left with a white
dwarf with $I_\mathrm{WD}=k_{\rm WD}^2 M_\mathrm{WD}
R_\mathrm{WD}^2$. If we now assume that $I_\mathrm{core} \simeq
I_\mathrm{WD}$ and ignore differences in radius of gyration,
conservation of angular momentum yields $\Omega_\mathrm{rot} /
\Omega_\mathrm{orb} \simeq 1+ M_\mathrm{env} R_\mathrm{L}^2 /
M_\mathrm{WD} R_\mathrm{WD}^2$. In reality, likely the envelope will
be more centrally concentrated than the white dwarf, i.e.,
$k_\mathrm{env}<k_{\rm WD}$, and tidal dissipation will be important
in the initial stages of the contraction. This will reduce the
spin-up. On the other hand, the hot core of the progenitor will be
larger than the white dwarf, i.e., $I_{\rm core}>I_{\rm WD}$. In any
case, it follows that unless the envelope mass is very small, the
white dwarf should be significantly spun up.

Model predictions for the envelope mass of helium-core white dwarfs
differ. The $0.196$\,M$_\odot$ model by \citet{sarb01}, has an
envelope mass of $6.7\times10^{-3}$\,M$_\odot$ (as given in
\citealt{asb01}), whereas a model of similar mass
($M_\mathrm{WD}=0.195$\,M$_\odot$) by \citet{dsbh98} has one of
$3.1\times10^{-2}$\,M$_\odot$. Using these values, taking
$M_\mathrm{WD}=0.19$\,M$_\odot$, $R_\mathrm{WD}=0.021$\,R$_\odot$ and
$R_\mathrm{L}=0.48$\,R$_\odot$, and ignoring differences in $k$, we
find current rotation periods a factor 18--85 faster than the
orbital period, or 20 to 5 minutes. Given that thick envelopes seem
inconsistent with the low observed temperature
(Sect.~\ref{sec:tandr}), the slower end of the range seems more
likely.

To estimate the timescale on which the pulsar flux is reprocessed, we
assume that the incident particles are predominantly highly energetic,
and that they penetrate to, roughly, one Thompson optical depth. This
corresponds to a column depth of $N=1.5\times10^{24}{\rm\,cm^{-2}}$,
for which the thermal timescale $t\simeq NkT/\sigma
T_\mathrm{eff}^4\simeq1\,$min, where the numerical estimate is for
$T=T_\mathrm{eff}=3700\,$K. This is shorter than the rotation periods
estimated above, suggesting that rotation may not be too important.
On the other hand, our estimate is very rough. For instance, at one
Thompson depth, the opacity at optical wavelengths is much smaller
than unity for the cool temperatures under consideration
\citep{sbl+94}. Thus, the material likely radiates less efficiently
than a black body, which would make the thermal timescale longer.
Furthermore, the irradiation will change the temperature and
ionisation structure of the atmosphere, further complicating matters.
(Indeed, could this be the underlying cause for the fact that the
colours deviate so strongly from those expected for a pure hydrogen
atmosphere?) Finally, it might induce strong winds which equalise the
temperature on both hemispheres (as is the case for Jupiter).

\section{Conclusions}\label{sec:discussion}
We have optically identified the white dwarf companion of the binary
millisecond pulsar PSR~J0751+1807. We find that the companion has the
reddest colours of all known millisecond pulsar companions and white
dwarfs. These colours indicate that the companion has a very low
(ultra-cool) temperature of
$T_\mathrm{eff}\sim\!3500-4300$\,K. Furthermore, the colours suggest
that the white dwarf has a pure helium atmosphere, or a helium
atmosphere with some hydrogen mixed in, as invoked for the field white
dwarf WD~0346+246 which has similar colours \citep{osh+01,ber01}. 

Our observations are inconsistent with evolutionary models, from which
one would expect a pure hydrogen atmosphere. Indeed, as for other
short-period systems, the hydrogen envelope is expected to be thick
enough to sustain significant residual hydrogen burning, leading to
temperatures far in excess of those observed. It may be that the mass
of the envelope was reduced due to shell flashes or irradiation by the
pulsar, as was proposed by \citet{esa01}.

However, we see no evidence for irradiation, despite the fact that the
pulsar spin-down flux inpinging on the white dwarf is roughly double
the observed thermal flux. Clues to what happens might be found from
more detailed studies of the spectral energy distribution, or more
accurate phase-resolved photometry.

Finally, a deeper observation at infrared wavelengths would allow one
to distinguish between the different atmosphere compositions for the
companion: for a pure helium atmosphere, black-body like colours are
expected, while if any hydrogen is present, the infrared flux would be
strongly depressed (as is seen for WD 0346+246). With adaptive optics
instruments, such observations should be feasible.

\begin{acknowledgements}
We thank Norbert Zacharias for providing preliminary UCAC2 data. We
also would like to thank the referee, Pierre Bergeron, for his useful
suggestions and for pointing out the existence of his updated
models. The observations for this paper were taken at the W. M. Keck
Observatory, which is operated by the California Association for
Research in Astronomy, a scientific partnership among the California
Institute of Technology, the University of California, and the
National Aeronautics and Space Administration. It was made possible by
the generous financial support of the W. M. Keck Foundation. MIDAS is
developed and maintained by the European Southern Observatory.  This
research made use of the SIMBAD and ADS data bases and of data
products from the Two Micron All Sky Survey, which is a joint project
of the University of Massachusetts and the Infrared Processing and
Analysis Center/California Institute of Technology, funded by the
National Aeronautics and Space Administration and the National Science
Foundation. We acknowledge support from NWO (C. G. B.), NSERC
(M. H. v. K.), and from NASA and NSF (S. R. K.).
\end{acknowledgements}

\bibliographystyle{aa}

\end{document}